\documentclass{article}
\usepackage{fortschritte}

\usepackage{latexsym}
\usepackage{amsmath}
\usepackage{amscd}
\usepackage{amsfonts}
\usepackage{amsthm}
\def\beq{\begin{equation}}                     %
\def\eeq{\end{equation}}                       %
\def\bea{\begin{eqnarray}}                     
\def\eea{\end{eqnarray}}                       
\def\tr{\mbox{tr}\,}                %
\begin {document}                 
\def\titleline{
%
%
%
Towards an Effective Action for D-Branes      
}
\def\email_speaker{
{\tt 
%
%
Alexandre.Sevrin@vub.ac.be            
}}
\def\authors{
%
%
%
%
%
A.~Keurentjes\1ad , P.~Koerber\2ad ,           
S.~Nevens\1ad , A.~Sevrin\1ad\sp, A.~Wijns\1ad 
}
\def\addresses{
%
%
%
%
\1ad   
Theoretische Natuurkunde, Vrije Universiteit Brussel and The International Solvay Institutes,\\
Pleinlaan 2, B-1050 Brussels, Belgium\\                                      %
\2ad                                        
University of British Columbia, Department of Physics and Astronomy,\\
6224 Agricultural Rd., Vancouver, B.C., V6T 1Z1, Canada           
}
\def\abstracttext{
%
%
We introduce and review several indirect methods to calculate the effective action for a single 
D-brane or a set of coinciding D-branes.
}
\large
\makefront
\section{Introduction}
The knowledge of the effective action for one or more D-branes is one of the few tools 
available to study D-brane dynamics. 
The effective world volume action for $n$ coinciding Dp-branes is, in leading order in
$ \alpha '$, given by the 
$d=9+1$, $N=1$ supersymmetric $U(n)$ Yang-Mills action dimensionally reduced to $p+1$ 
dimensions \cite{witten}. For a single D-brane, $n=1$, the effective action is known to all
orders in $ \alpha '$ in the limit of constant (or slowly varying) background fields: it is the
$d=9+1$, $N=1$ supersymmetric Born-Infeld action, dimensionally reduced to $p+1$ 
dimensions, \cite{BI1} -- \cite{BI7}. Both bosonic and fermionic terms as well as the couplings to the 
bulk background fields are known. Derivative corrections were studied in 
\cite{andreevtseytlin} (using the partition function method), in \cite{wyllard} (using boundary
conformal field theory) and \cite{cornalba} (using the Seiberg-Witten map). Modulo field 
redefinitions, it was shown that there are no two derivative corrections and a proposal for the four 
derivative corrections was made through all orders in $ \alpha '$.

For $n>1$, the situation is more involved. Requiring the background fields to be constant brings one, 
because of $D_aF_{bc}=0\Rightarrow [F_{ab},F_{cd}]=0$, back to the abelian situation. So here, one has 
to deal with derivative corrections right from the start. Only partial results are known. Indeed,
there are no
$ {\cal O}( \alpha ')$ corrections and the $ {\cal O}( \alpha '{}^2)$ corrections were calculated from
open superstring amplitudes in \cite{direct}. Requiring that certain BPS configurations solve the 
equations of motion allowed one to calculate both the $ \alpha '{}^3$ \cite{alpha3}, and the 
$ \alpha '{}^4$ \cite{alpha4}, corrections (see also the summarizing equations in 
\cite{testalpha4}). 

As a direct calculation starting from open string scattering amplitudes is technically very involved, 
indirect methods are called for. In this paper we will review some of those and outline
some of the future strategies.

\section{Linear Constraints on Magnetic Fields}
One of the most powerful methods to construct the effective action for D-branes rests on the 
requirement that certain configurations which generalize $d=4$ instantons solve the equations of 
motion. In this section we review these configurations in the $ \alpha '\rightarrow 0$ limit.

The Yang-Mills equations of motion\footnote{We
work in flat Euclidean space of dimension $d=2m$. Unless stated otherwise, we sum over repeated 
indices regardless their position. In the next sections we will often put $2 \pi \alpha '=1$.},
\begin{eqnarray}
D_bF_{ba}=0,
\end{eqnarray}
are solved by virtue of the Bianchi identities if one imposes,
\begin{eqnarray}
F_{ab}\propto \varepsilon _{abcd}F_{cd},
\end{eqnarray}
where $ \varepsilon _{abcd}$ is completely anti-symmetric in its indices. If one insists on preserving
Lorentz covariancy, one is limited to $d=4$ and one deals with instantons. In \cite{Corrigan:1982th} 
the maximal subgroups of $SO(2m)$ for which $ \varepsilon $ transforms as a scalar were catalogued. 
Two 
generic cases emerged: $U(m)\subset SO(2m)$ and $SO_+(7)\subset SO(8)$. The latter is one of the 
$SO(7)$ subgroups of $SO(8)$ for which the vector and one of the spinor representations decompose as 
$8=8$, while the other spinor representation decomposes as $8= 1\oplus 7$. This result is clarified if 
one passes to complex coordinates. The equations of motion become,
\begin{eqnarray}
0&=&D_{ \beta }F_{ \bar \beta \alpha }+ D_{ \bar \beta }F_{  \beta \alpha } \nonumber\\
&=&D_{ \alpha }F_{ \bar \beta \beta  }+ 2D_{ \bar \beta }F_{  \beta \alpha },
\end{eqnarray}
where we used the Bianchi identities. The equations of motions are solved provided one imposes,
\begin{eqnarray}
F_{ \alpha \beta }= F_{ \bar \alpha \bar \beta }=0,\qquad F_{ \alpha \bar \alpha }=0,\label{shb1}
\end{eqnarray}
where in the last equation a summation over $ \alpha $ is understood.
This corresponds to the case $SU(m)\subset SO(2m)$. The first two equations are recognized as 
holomorphicity conditions while the last is the stability condition; eq.~(\ref{shb1}) defines a stable 
holomorphic bundle. Passing back to real coordinates, one verifies that for $m=2$,
these are equivalent to the instanton equations. The exceptional case $SO_+(7)\subset SO(8)$ is also 
easily dealt with. Indeed taking $m=4$ and modifying eq.~(\ref{shb1}) to,
\begin{eqnarray}
F_{ \alpha \beta }= \frac 1 2 \varepsilon _{ \alpha \beta \gamma \delta }F_{ \bar \gamma  
\bar \delta  },\qquad F_{ \alpha \bar \alpha }=0,\label{defoi}
\end{eqnarray}
one again solves the equations of motion. These equations 
are known as the octonionic instanton equations which generalize the ordinary -- also known as 
quaternionic -- instanton equations. 
Turning back to real coordinates, eq.~(\ref{defoi}) can be 
rewritten as, 
\begin{eqnarray}
F_{8a}=\frac 1 2 f_{abc}F_{bc},\qquad a,\,b,\,c\,\in\{1,\cdots ,7\},\label{defoi1}
\end{eqnarray}
with $ f_{abc}$ the (completely anti-symmetric) octonionic structure constants,
\begin{eqnarray}
f_{127}=f_{163}=f_{154}=f_{253}=f_{246}=f_{347}=f_{567}=1.\label{osc}
\end{eqnarray}
Written in this way, it is clear that this is the $d=8$ generalization of the well known
$d=4$ (quaternionic) instantons,
\begin{eqnarray}
F_{4a}=\frac 1 2 \varepsilon _{abc}F_{bc},\qquad a,\,b,\,c\,\in\{1,2,3\}.\label{d4bps}
\end{eqnarray}

For $m\leq 4$, the solutions discussed above are  BPS configurations of D-branes. For generic
values of the magnetic fields satisfying eq.~(\ref{shb1}), one has 8, 4, 2 unbroken supersymmetries
resp.~for $m=2$, $m=3$, $m=4$ resp. The octonionic configuration, eq.~(\ref{defoi}) leaves a single 
supersymmetry unbroken. Explicit solutions to these equations can be interpreted as certain stable 
arrangements of D-branes. An exception on this are solutions of the 
octonionic type which generically have no concrete realization in terms of D-branes. 
However, even there some progress is being made. Indeed, starting e.g. from 
eq.~(\ref{d4bps}) in the abelian limit and dimensionally reducing it to $d=3$, one obtains the Dirac 
monopole which can be viewed as a single D3-brane with a stack of D1-branes perpendicular to it. The 
number of D1-branes corresponds to the monopole number. This is the so-called BIon 
configuration \cite{Callan:1997kz}, \cite{Gibbons:1997xz}. Similarly,  eq.~(\ref{defoi}) in the 
abelian limit dimensionally reduced to $d=7$ can be reinterpreted
as a single D7-brane with a configuration of D5-branes each having 4 directions longitudinal to the 
D7-brane. In addition several stacks of D3-branes and a stack of D1-branes can be added 
\cite{usinprep1}. Their possible locations are determined by the octonionic structure constants.

Note that the BPS equations for $m>2$ do get $ \alpha '$ corrections. As these solutions have to 
solve the equations of motion of the effective action, it was suggested in \cite{Denef:2000rj} that 
they might provide a handle on the $ \alpha '$ corrections in the effective action. In the next 
section we will see that this is indeed the case.
\section{The Effective Action from Stable Holomorphic bundles}
A natural question which arises is whether we can deform the Yang-Mills
action in such a way that stable holomorphic bundles -- or some deformation of it -- remain solutions  
to the equations of motion. As an example, we look at the abelian case and assume the fieldstrengths 
to be constant. We then add terms polynomial in the field strength to the action ignoring derivative 
terms, arriving at the following most general
action through order $\alpha'{}^2$,
\begin{equation}
{\cal S}=
\frac{1}{g_{YM}^2}\int\left(\frac{1}{4} \tr F^2 + \lambda_{1} \tr F^4 + \lambda_{2} \left(\tr 
F^2\right)^2 + {\cal O}\left(\alpha'{}^4 \right)\right) \, ,
\label{lagrangian2}
\end{equation}
with $\lambda_{1}$ and $\lambda_{2}$ arbitrary real coefficients\footnote{$F^l{}_{ab}$ stands for 
$F_{ac_1}F_{c_1c_2}\cdots
F_{c_{l-1}b}$ and $\tr F^l\equiv F^l{}_{aa}$. }. {From} this the equations of motion 
follow,
\begin{eqnarray}
0&=&\partial_{\bar{\beta}} \left( F_{\alpha\bar{\alpha}} + \frac{8 \lambda_{1}}{3} 
F^3_{\alpha\bar{\alpha}}\right)+ \left( 4 \lambda_{1} + 16 \lambda_{2} \right) 
\partial_{\bar{\gamma}} F^2_{\alpha\bar{\alpha}} F_{\gamma\bar{\beta}}
+ \nonumber\\
&&16 \lambda_{2} F^2_{\gamma\bar{\gamma}} \partial_{\bar{\beta}} F_{\alpha\bar{\alpha}}
+ 8 \lambda_{1} F^2_{\gamma\bar{\beta}} \partial_{\bar{\gamma}} F_{\alpha\bar{\alpha}} + {\cal 
O}(\alpha'{}^4 ),
\label{eom2}
\end{eqnarray}
where we passed to complex coordinates, used the Bianchi identities and implemented the holomorphicity
conditions. The first term vanishes provided we deform the stability condition,
\begin{equation}
F_{\alpha \bar\alpha}+\frac{8\lambda_{1}}{3} F^3_{\alpha\bar\alpha} + 
{\cal O}(\alpha'{}^4)=0 \, ,
\label{duy1}
\end{equation}
while the second term vanishes if,
\begin{equation}
\lambda_{2} = -\frac{1}{4} \lambda_{1} \, .
\label{cond2}
\end{equation}
The stability condition in leading order takes care that the remainder vanishes as well, but because 
of eq.~(\ref{duy1}) they will make a contribution at order $\alpha'{}^4$ thereby relating 
$\lambda_{1}$ to the arbitrary coefficients at that order. 

We see that at this point there are two undetermined coefficients left:
an overall multiplicative constant $1/g_{YM}^2$
and $\lambda_{1}$. The latter can be absorbed by rescaling $F$, putting 
$ \lambda _1=1/8$, its conventional value.
Proceeding to higher orders, one finds that all coefficients get uniquely fixed, yielding
the Born-Infeld action \cite{DeFosse:2001mk}!

This method can be extended to the non-abelian case as well. In this way the action was obtained
through order $ \alpha '{}^4$ \cite{alpha3}, \cite{alpha4}. 
Modulo field redefinitions it is given by,
\begin{eqnarray}
{\cal L}=\frac{1}{g^2}\left({\cal L}_0+{\cal L}_2+{\cal L}_3+{\cal L}_4\right),
\end{eqnarray}
where the leading term is simply\footnote{
The $u(n)$ generators are taken to be hermitian. We always trace over the Lorentz indices and Tr and 
STr resp.~stand for the group trace and the symmetrized group trace. When symmetrizing, we take $D^l 
F$ as a single unit.}
\begin{eqnarray}
\label{la0}
{\cal L}_0=\;=\; -
 \mbox{Tr}\, \left\{\frac{1}{4}F^2\right\}\,.
\end{eqnarray}
Subsequently we have
\begin{eqnarray}
\label{la2}
{\cal L}_2=\mbox{STr} \left\{\frac{1}{8} F^4
- \frac{1}{32} F^2F^2
\right\},
\end{eqnarray}
and
\begin{eqnarray}
\label{la3}
{\cal L}_3=\frac{\zeta (3)}{2\pi^3}\mbox{Tr}\left\{[D_3,D_2] D_4 F_{51} \, D_5 [D_4,D_3] F_{12} 
\right\}.
\end{eqnarray}
The overall coefficient of this term remained undetermined\footnote{This is fortunate, as
our method can only generate rational coefficients while $\zeta (3)/ \pi ^3$ is very
probably an irrational number.} when
using the method of \cite{DeFosse:2001mk}. It was fixed by comparing it to
the partial result for this term in \cite{bilal1} 
obtained by a direct string theoretic calculation. Finally the fourth order term is completely
determined by the method of \cite{DeFosse:2001mk} and it is given by,
\begin{eqnarray}
{\cal L}_4 = {\cal L}_{4,0} + {\cal L}_{4,2} + {\cal L}_{4,4} \, ,\label{la4}
\end{eqnarray}
with
\begin{equation}
\begin{split}
{\cal L}_{4,0} & =  -\mbox{STr} \left( \frac{1}{12} F_{12}F_{23}F_{34}F_{45}F_{56}F_{61}
                 - \frac{1}{32} F_{12}F_{23}F_{34}F_{41}F_{56}F_{65}
                  + \frac{1}{384} F_{12}F_{21}F_{34}F_{43}F_{56}F_{65} \right) , \\
{\cal L}_{4,2} & = -\frac{1}{48} \mbox{STr} \Big( -2 \, F_{12}D_{1}D_{6}D_{5}F_{23}D_{6}F_{34}F_{45}
                                     - F_{12}D_{5}D_{6}F_{23}D_{6}D_{1}F_{34}F_{45}  \\
               & + 2 \, F_{12}\left[D_{6},D_{1}\right] D_{5}F_{23}F_{34}D_{4}F_{56}
                 + 3 \, D_{4}D_{5}F_{12}F_{23}\left[D_{6},D_{1}\right]F_{34}F_{56} \\
               & + 2 \, D_{6}\left[D_{4},D_{5}\right]F_{12}F_{23}D_{1}F_{34}F_{56}
                 + 2 \, D_{6}D_{5}F_{12}\left[D_{6},D_{1}\right]F_{23}F_{34}F_{45} \\
               & + 2 \, \left[D_{6},D_{1}\right]D_{3}D_{4}F_{12}F_{23}F_{45}F_{56} \\
               & + \left[D_{6},D_{4}\right]F_{12}F_{23}\left[D_{3},D_{1}\right]F_{45}F_{56} \Big) , \\
{\cal L}_{4,4} & = -\frac{1}{1440} \mbox{STr} \Big( D_6 [D_4,D_2]D_5 D_5 [D_1,D_3] D_6 F_{12} F_{34}
                   + 4 \, D_2 D_6 [D_4,D_1][D_5,[D_6,D_3]] D_5 F_{12} F_{34}  \\
                 & + 2 \, D_2 [D_6,D_4][D_6,D_1] D_5 [D_5,D_3] F_{12} F_{34}
                   + 6 \, D_2 [D_6,D_4]D_5[D_6,D_1][D_5,D_3] F_{12} F_{34} \\
                 & + 4 \, D_6 D_5 [D_6,D_4][D_5,D_1][D_4,D_3] F_{12} F_{23}
                   + 4 \, D_6 D_5 [D_4,D_2][D_6,D_1][D_5,D_3] F_{12} F_{34} \\
                 & + 4 \, D_6 [D_5,D_4][D_3,D_2][D_5,[D_6,D_1]] F_{12} F_{34} \\
                 & + 2 \, [D_6,D_1][D_2,D_6][D_5,D_4][D_5,D_3]F_{12}F_{34} \Big) \, .
\end{split}\label{la4e}
\end{equation}

It is clear that the result is quite complicated. In this form it does not particularly suggest any 
all order expression. The lesson which can be learned here is that it might be a good idea to return 
to the simpler abelian case and try to obtain control over the derivative corrections. Precisely this 
will be tackled in the next section. 

\section{The Effective Action from $ \beta $-Functions}
In this section we turn to an alternative way to recover the effective action. We consider
an open string $ \sigma $-model in a $U(1)$ background and require the model to be
conformally invariant at the quantum level, i.e. we require the $ \beta $-functions to vanish which
is viewed as an equation of motion which subsequently must be integrated to the effective action. 
The $n$-loop contribution to the $ \beta $-functions gives $2n-2$-derivative corrections to the 
effective action. A great advantage of this approach is that the results are always all order in
$ \alpha '$. At lowest order, this was already studied in \cite{BI2}, see also \cite{Klaus}. The
calculations are greatly simplified when formulating the model in $N=2$ boundary
superspace. In this setting the whole $U(1)$ structure is characterized by a single
scalar potential $V$. The model at hand is a special case of the general setup developed
in \cite{Koerber:2003ef}.

We introduce chiral fields, $Z^ \alpha $, and anti-chiral fields , $Z^{\bar \alpha }$, $ \alpha 
\in\{1,\cdots m\}$, satisfying the constraints,
\begin{eqnarray}
\bar D Z^ \alpha =D Z^{\bar \alpha }=0.\label{con1}
\end{eqnarray}
In addition we need a set of fermionic constrained fields $ \Psi^ \alpha  $ and $ \Psi ^{\bar 
\alpha }$, which satisfy,
\begin{eqnarray}
\bar D \Psi ^ \alpha = \partial _ \sigma  Z^ \alpha ,\qquad
D \Psi ^{\bar \alpha }= \partial _ \sigma Z^{\bar \alpha }.\label{con2}
\end{eqnarray}
The action, $ {\cal S}$, consists of a free bulk term, $ {\cal S}_0$, 
and a boundary interaction term, $ {\cal S}_{int}$,
\begin{eqnarray}
{\cal S}= {\cal S}_0+ {\cal S}_{int},
\end{eqnarray}
where,
\begin{eqnarray}
{\cal S}_0&=&  \int d^2 \sigma d^2 \theta \left(
g_{ \alpha \bar \beta }DZ^ \alpha\bar DZ^{\bar \beta }+
g_{ \alpha \bar \beta } \Psi ^ \alpha \Psi ^{\bar \beta }  
\right), \nonumber\\
{\cal S}_{int}&=&-\int d \tau d^2 \theta V\left(
Z, \bar Z\right).
\end{eqnarray}
In these equations, we rescaled $Z$ ($ \Psi $ resp.) by a factor 
$\sqrt{2 \pi  \alpha '}$ such as to make 
it dimensionless (of dimension $1/2$ resp.). We choose Neumann boundary conditions,
\begin{eqnarray}
\Psi^ \alpha  \Big|{}_{\mbox{boundary}}=\Psi^{\bar \alpha } \Big|{}_{\mbox{boundary}}=0.
\label{bdycds}
\end{eqnarray} 
The boundary term gives the coupling to a $U(1)$ background field. The magnetic fields, appropriately 
rescaled by a factor $2 \pi  \alpha '$, are obtained from the potential,
\begin{eqnarray}
&&F_{ \alpha \bar \beta }= i V_{ \alpha \bar \beta } \nonumber\\
&&F_{ \alpha \beta }=F_{\bar \alpha \bar \beta }=0.\label{fspot}
\end{eqnarray}

Calculating and resumming the UV divergent one loop contributions yields, using minimal
subtraction, a counterterm,
\begin{eqnarray}
V_{bare}=V+i\,\frac{ \ln\, \mu}{   \pi}  \, g^{ \alpha \bar \beta }
\left(\mbox{arcth}\,F\right)_{ \alpha \bar \beta },
\end{eqnarray}
with $ \mu $ the UV cut-off. Requiring the one-loop $ \beta $-function to vanish yields,
\begin{eqnarray}
{\cal G}^{bc} \partial _b F_{ca}=0,\label{bieom}
\end{eqnarray}
where $ {\cal G}$ is defined by,
\begin{eqnarray}
{\cal G}^{ac}(g_{cb}-(F^2)_{cb})= \delta ^a_b.
\end{eqnarray}
We immediately recognize eq.~(\ref{bieom}) as the equation of motion for the Born-Infeld action. 
Proceeding to the
next order one finds that the $\ln \,\mu $ divergences cancel between the two-loop diagrams and the 
subdivergent diagram, leaving a $(\ln\, \mu )^2$ divergence. As a consequence, the $ \beta $-function 
does 
not receive a two-loop contribution. Put differently: there are no two-derivative terms in the 
effective action, a result already derived in \cite{andreevtseytlin}. Presently the 3- and 4-loop 
contributions are being calculated and all order properties are being investigated \cite{usinprep2}.

\section{Conclusions}
It is clear that an all order expression for the effective action for D-branes will not be available
in an immediate future. However, combining various indirect methods such as deformed stable 
holomorphic bundles and $ \sigma $-model $\beta $-functions will probably lead to concrete results
for at least the abelian case.

{\bf Acknowledgement}  
We thank Klaus Behrndt, Marc Grisaru and Chris Hull for useful discussions.
SN and AW are ``FWO aspiranten''.
AK, SN, AS and AW are supported in part by the ``FWO-Vlaanderen''
through project G.0034.02, in part by the Belgian Federal Science Policy Office 
through the
Interuniversity Attraction Pole P5/27 and in part by the European
Commission FP6 RTN programme MRTN-CT-2004-005104. PK is supported by the Natural Sciences and 
Engineering Research Council of Canada.


\end{document}